\documentclass[12pt]{iopart}
\usepackage[scr=boondoxo,scrscaled=1.05]{mathalfa}
\usepackage{tikz}
\usetikzlibrary{calc,decorations.markings}
\usepackage{epsfig}
\usepackage{graphicx}
\usepackage{rotating}
\expandafter\let\csname equation*\endcsname\relax
\expandafter\let\csname endequation*\endcsname\relax
\usepackage{amsmath,amssymb}
\usepackage[a4paper]{geometry}
\usepackage{subcaption}
\usepackage{float} 
\usepackage[font=small,labelfont=bf,justification=justified,format=plain]{caption}
\usepackage[active]{srcltx}
\usepackage{url}
\usepackage{centernot}
\usepackage{dsfont,bm}
\usepackage{pgfplots}
\usepackage{stackengine}
\usepackage{enumitem}
\usepackage[scr=boondoxo,scrscaled=1.05]{mathalfa}
\usepackage{hyperref}
\colorlet{darkred}{red!85!black}
\colorlet{darkgreen}{green!50!black}
\colorlet{darkblue}{blue!60!black}
\hypersetup{
	colorlinks   = true, 
	urlcolor     = darkgreen, 
	linkcolor    = darkblue, 
	citecolor   = darkred 
}
\usepackage{tikz}
\usepackage{tkz-euclide}
\usetikzlibrary{calc,decorations.markings}
\usepackage{times,graphicx,xcolor}
\usepackage{delarray,fancybox}
\usepackage{mathdots}
\usepackage{eurosym}
\usepackage{relsize}
\usepackage{soul}

\colorlet{darkred}{red!85!black}
\colorlet{darkgreen}{green!50!black}
\colorlet{darkblue}{blue!60!black}
\definecolor{newred}{RGB}{125,0,45}
\definecolor{newblue}{RGB}{0,80,158}
\definecolor{newgray}{gray}{0.4}
\definecolor{ArmyGreen}{rgb}{0.29, 0.33, 0.13}
\definecolor{BostonRed}{rgb}{0.9, 0.0, 0.0}
\definecolor{CadmiumGreen}{rgb}{0.0, 0.42, 0.24}
\definecolor{darkyellow}{RGB}{205,205,0}
\definecolor{Gray}{gray}{0.9}
\definecolor{LightCyan}{rgb}{0.88,1,1}
\definecolor{PastelGreen}{rgb}{0.47,0.87,0.47}
\definecolor{PastelRed}{rgb}{0.87,0.47,0.47}
\definecolor{PastelYellow}{rgb}{0.99, 0.99, 0.59}
\definecolor{platinum}{rgb}{0.9, 0.89, 0.89}
\definecolor{pistachio}{rgb}{0.58, 0.77, 0.45}
\definecolor{princetonorange}{rgb}{1.0, 0.56, 0.0}
\definecolor{FigureBlue}{rgb}{0.447, 0.624, 0.812}
\definecolor{NewBlue}{rgb}{0.712, 0.806, 0.861}
\definecolor{LightBlue}{rgb}{0.812, 0.906, 0.961}
\definecolor{airforceblue}{rgb}{0.36, 0.54, 0.66}
\definecolor{amber}{rgb}{1.0, 0.75, 0.0}

\def\<{\langle}
\def\>{\rangle}

\newcommand{\td}[1]{{\scalebox{0.8}{\ensuremath{(#1)}}}}

\begin{document}
	
	\title[On the use of the Belopol'skaya-Daletskii representation of a diffusion]{On the use of the Belopol'skaya-Daletskii representation of a diffusion on a Riemann manifold to construct path integrals}
	
	\author{Paolo Muratore-Ginanneschi}
	
	\address{University of Helsinki, Department of Mathematics and Statistics
		P.O. Box 68 FIN-00014, Helsinki, Finland. \\
	} 
	\vspace{-0.2cm}
	\ead{paolo.muratore-ginanneschi@helsinki.fi}
\begin{abstract}
We show that the Belopol'skaya-Daletskii formulation of stochastic differential equations on a Riemann manifold offers an elementary way to construct equivariant representations of finite-dimensional approximations to the path measure of a diffusion. 
The key ingredient is the use of the exponential map to describe increments of the diffusion. 
\end{abstract}

\section{Introduction}

Diffusion processes generated by a state-dependent noise, often referred to as multiplicative noise, are central to the investigation of many areas of non-equilibrium statistical physics. For instance, multiplicative noise models come about from first-principle analytic derivations of open system dynamics of a classical {\textquotedblleft}central{\textquotedblright} particle interacting with multiple baths at different temperatures and subject  to local friction \cite{vKaN1988,JaMa1995}.  
State-dependent noise 
may significantly affect the relative stability and induce transport in open systems otherwise at equilibrium in the presence of purely thermal noise \cite{ButM1987b,LanR1988}. For these reasons, the statistics of the entropy production derived from fluctuation relations  \cite{ChGa2008,DaCoPa2023}  and the conditions under which a Gibbs measure is reversible in the presence of state-dependent noise continue to attract considerable attention in stochastic thermodynamics \cite{OByCa2024,AyDiPaZi2025,LuBaViVu2025}. Recently even more so, because machine learning techniques guided by Rayleigh-Onsager variational principles \cite{MieA2011,BeDeGaJoLa2015}  are used to reconstruct the macroscopic dynamics of systems consisting of a large yet finite number of particles in the form of stochastic (partial) differential equations with multiplicative noise, see e.g., \cite{HuHeDiZiRe2025}. Mathematically rigorous constructions of solutions of stochastic partial differential equations for the macroscopic dynamics are also an active field of research \cite{BrGaHaZa2021}. Finally, stochastic differential equations with state-dependent noise describe fluid particle separation dynamics in the theory of fluid turbulence \cite{FaGaVe2001,GaHo2004}  where they are the basis of Monte Carlo methods for numerically generating  the statistics of clusters of fluid particles, see e.g., \cite{FrMaVe1998,MaMG2000}.

The above considerations motivate the continued interest  (see, for instance, \cite{ItoH1978} and the recent \cite{CuLe2017,PiCuLeWi2023,DioL2024}) in a geometrically transparent characterization of stochastic differential equations with multiplicative noise. Riemann geometry plays a role even when the diffusion process takes values on a standard $d$-dimensional Euclidean vector space. The reason is that a local, non-singular, diffusion tensor induces a Riemannian structure in which the metric is specified by its inverse.   At the same time,  the semi-martingale decomposition of the process in generic local coordinates specifies a drift field in the It\^o representation of the stochastic differential equation that does not transform as a (contravariant) vector field in consequence of the finite quadratic variation of the Wiener process.  
This fact was already pointed out  by Kyosi It\^o at the dawn of the theory of stochastic differential equations \cite{ItoK1950}. 
The definition of stochastic differential equations with multiplicative noise depends on the discretization prescription. One may think that is possible to describe the drift as a vector field by representing the equation in the Stratonovich ({\textquotedblleft}mid-point{\textquotedblright}) 
discretization. 
Unfortunately, this is not the case. Citing \cite{BrGaHaZa2021}, the tenet is that {\textquotedblleft}\emph{in the classical case of finite-dimensional stochastic differential equations there does not exist any natural notion of solution which is equivariant under changes of coordinates and satisfies It\^o’s isometry simultaneously}{\textquotedblright}. In other words, and still citing \cite{BrGaHaZa2021}, {\textquotedblleft}\emph{there does not appear to exist any notion of stochastic integration $\star$ which is defined on a natural class of integrands and such that the process solves}{\textquotedblright}
		\begin{align*}
			\mathrm{d}\bm{\mathscr{q}}\td{t}=\bm{h}(\bm{\mathscr{q}}\td{t})\mathrm{d}t+\mathsf{A}(\bm{\mathscr{q}}\td{t})\star\mathrm{d}\bm{\mathscr{w}}\td{t}
		\end{align*}
where $\bm{h}$ and $ \mathsf{A}\star\mathrm{d}\bm{\mathscr{w}}\td{t}$ transform as contravariant vector fields and $\bm{\mathscr{w}}$ is a $d$-dimensional Wiener process  (see also \cite[\S~4]{ItoH1978}).

The lack of an equivariant representation of the stochastic differential equations is highly consequential for the construction of the path measure from finite-dimensional approximations. Elementary derivations based on the stochastic differential equation and the pre-point discretization necessarily involve Christoffel symbols \cite{GraR1985}. Rigorous and technically laborious constructions based on asymptotic analysis of the Fokker-Planck equation \cite{InMa1985} or on the lattice limit \cite{AnDr1999,BaPf2008} have the great merit of proving the existence of a covariant path integral representation of the scalar transition density. They do not provide, however, an explicit relation with the more elementary and computationally convenient derivation \cite{GraR1985}. Furthermore, \cite[Th~6.1]{InMa1985} and \cite[Lemma~4.3]{BaPf2008} prove that depending on the choice of measure on the approximating spaces of geodesic polygons one can modify the action functional to include a term proportional to the scalar curvature of the metric and the numerical value of the prefactor determined by the choice of  measure. This is not surprising: in the presence of multiplicative noise the same diffusion process satisfies infinitely many equivalent stochastic differential equations. Because the proofs in \cite{AnDr1999,BaPf2008} are highly technical, the implications of these papers, despite their great merits, are not immediate for a reader interested in explicit rules of calculus for performing perturbative or numerical computations based on the expression of the path integral.

The scope of this contribution is to show how a representation of It\^o stochastic differential equations on Riemann manifolds introduced long ago by Belopol'skaya and Daletskii (alternative transliteration Belopolskaya and Dalecky) \cite{BeDa1982}, see also \cite{BeDa1990,GliY1996}, provides a transparent way to reconcile the elementary approach of \cite{GraR1985} with the rigorous results of \cite{InMa1985,AnDr1999,BaPf2008}. We proceed heuristically, without claiming full rigor. We only avail ourselves of elementary algebra and the well-known properties of the quadratic variation of the Wiener process. We thus expect that our line of reasoning can also be turned into a rigorous proof. After briefly recalling in section~\ref{sec:pre} some basic concepts of stochastic differential geometry which can be skipped by the familiar reader, in section~\ref{sec:BP} we present the basic ideas  inspiring the Belopol'skaya-Daletskii formalism. Next, in section~\ref{sec:pi} we apply the formalism to the construction of finite-dimensional approximations  to the path measure from the short-time asymptotics of the scalar density. This section contains the main contribution of this note.  In section~\ref{sec:exe}, we show how finite-dimensional approximations to the path measure recover the mean forward derivative on test scalar and vector fields i.e., the Dohrn-Guerra formula \cite{DoGu1978,DoGu1979} which governs the diffusion of vector and tensor fields on a Riemann manifold. 
The last section is devoted to conclusions.  

To simplify the treatment, we always assume that the Riemann manifold is $\mathbb{R}^{d}$ endowed with a non-trivial metric specified by a strictly non-vanishing  diffusion tensor.

\section{Mathematical preliminaries}
\label{sec:pre}

We characterize a diffusion process $\bm{\mathscr{q}}$ in $\mathbb{R}^{d}$ by means of its drift
\begin{equation}
	\label{pre:drift}
	\bm{b}(\bm{q})=\lim_{s\downarrow 0} \operatorname{E}\left(\frac{\bm{\mathscr{q}}\td{t+s}-\bm{\mathscr{q}}\td{t}}{s}\Big{|}\bm{\mathscr{q}}\td{t}=\bm{q}\right)
\end{equation}
and diffusion tensor
\begin{equation}
	\label{pre:D}
	\mathsf{D}(\bm{q})=\lim_{s\downarrow 0}\operatorname{E}\left(\frac{\left (\bm{\mathscr{q}}\td{t+s}-\bm{\mathscr{q}}\td{t}-\bm{b}(\bm{\mathscr{q}}\td{t})\,s\right )\otimes\left (\bm{\mathscr{q}}\td{t+s}-\bm{\mathscr{q}}\td{t}-\bm{b}(\bm{\mathscr{q}}\td{t})\,s\right )}{s}\Big{|}\bm{\mathscr{q}}\td{t}=\bm{q}\right)
\end{equation}
We assume here all needed regularity  for the drift field and the diffusion tensor $\mathsf{D}$ as functions of their argument, see e.g., \cite[\S~IV]{IkWa1989}. By definition, the diffusion tensor specifies a positive symmetric matrix for every $\bm{q}\in \mathbb{R}^{d}$. Furthermore, we suppose the diffusion tensor to be everywhere non-vanishing, i.e., it is uniformly positive.  Hence, there exists a complete basis of real eigenvectors $\big{\{}\bm{\mathrm{e}}_{i}(\bm{q})\big{\}}_{i=1}^{d}$ \emph{orthogonal} with respect to the standard inner product in $\mathbb{R}^{d}$ which reduces the diffusion tensor to diagonal form
\begin{align*}
	\mathsf{D}(\bm{q})=\sum_{i=1}^{d}\bm{\mathrm{e}}_{i}(\bm{q})\bm{\mathrm{e}}_{i}^{\top}(\bm{q})
\end{align*}
In writing this expression we use the fact that eigenvalues are positive and can be reabsorbed in the definition of the eigenvectors.
We refer to the collection of these eigenvectors as an orthogonal frame. Geometrically, we can interpret each of the $\bm{\mathrm{e}}_{i}(\bm{q})$  individually as a (contravariant) vector field \cite[\S~V]{IkWa1989}: we derive their transformation law under a diffeomorphism $$\bm{\Phi}\colon\mathbb{R}^{d}\mapsto\mathbb{R}^{d}$$ 
by applying to this latter the collection of scalar differential operators:
\begin{align*}
&	\mathfrak{E}_{i}=\bm{\mathrm{e}}_{i}^{k}(\bm{q})\partial_{k}&& i=1,\dots,d
\end{align*}
where, as customary, we use Einstein convention on repeated indices to denote contraction.

 Algebraically, $\bm{\mathrm{e}}_{i}$ specifies the $i$-th column of the volatility matrix $\mathsf{A}$:
\begin{align}
	\begin{split}
		&\mathsf{A}(\bm{q})=\begin{bmatrix}	\bm{\mathrm{e}}_{1} & \dots &\bm{\mathrm{e}}_{d} \end{bmatrix}
		\\
		&\mathsf{D}(\bm{q})=\mathsf{A}(\bm{q})\mathsf{A}^{\top}(\bm{q})
	\end{split}
	\label{pre:volatitity}
\end{align}
We identify the inverse of the diffusion tensor  with a Riemann metric tensor on $\mathbb{R}^{d}$:
\begin{align*}
	\mathsf{g}(\bm{q})=\mathsf{D}^{-1}(\bm{q})
\end{align*}
which we can use to map contravariant vectors into covariant ones. In particular, we find
\begin{align}
	\bm{\tilde{\mathrm{e}}}_{i}(\bm{q})=\mathsf{g}(\bm{q})\bm{\mathrm{e}}_{i}(\bm{q})
	\label{pre:cf}
\end{align} 
and
\begin{align*}
	\mathsf{g}(\bm{q})=\sum_{i=1}^{d}\bm{\tilde{\mathrm{e}}}_{i}(\bm{q})\bm{\tilde{\mathrm{e}}}_{i}^{\top}(\bm{q})
\end{align*}
This also means that elements of the frame are orthonormal with respect to the inner product specified by the Riemann metric
\begin{align*}
	\left \langle\,\bm{\mathrm{e}}_{i}(\bm{q})\,,\bm{\mathrm{e}}_{j}(\bm{q})\,\right\rangle_{\mathsf{g}}=\delta_{i,j}
\end{align*} 
Once we are equipped with a Riemann metric, we can construct a Levi-Civita connection by imposing the requirements of no torsion and metric compatibility:
\begin{align}
&0=\nabla_{i}\mathsf{g}_{\ell,k}=\partial_{i}\mathsf{g}_{\ell,k}-\Gamma^{j}_{i,\ell}\mathsf{g}_{j,k}-\Gamma^{j}_{i,k}\mathsf{g}_{\ell,j}&& \forall\,i,\ell,k=1,\dots,d
\label{pre:mcompatibility}
\end{align}
 where
\begin{align*}
	\Gamma^{\ell}_{i,j}=\Gamma^{\ell}_{j,i}=\frac{1}{2}\mathsf{D}^{\ell,k}\big{(}\partial_{i}\mathsf{g}_{k,j}+\partial_{j}\mathsf{g}_{k,i}-\partial_{k}\mathsf{g}_{i,j}\big{)}
\end{align*}
are the Christoffel symbols. From (\ref{pre:mcompatibility}), it also follows that
\begin{align*}
0=	\nabla_{i}\mathsf{D}^{\ell,k}=\partial_{i}\mathsf{D}^{\ell,k}+\Gamma^{\ell}_{i,j}\mathsf{D}^{j,k}+\Gamma^{k}_{i,j}\mathsf{D}^{\ell,j}&& \forall\,i,\ell,k=1,\dots,d
\end{align*}
It is also useful to recall how the connection acts on scalar and on the components of a contravariant vector and covariant fields \cite[\S~9]{FraT2012}:
\begin{align}
	\begin{split}
&	\nabla_{i}f=\partial_{i}f
\\
&\nabla_{i}v^{\ell}=\partial_{i}v^{\ell}+\Gamma^{\ell}_{i,j}v^{j}
\\
&\nabla_{i}v_{\ell}=\partial_{i}v_{\ell}-\Gamma^{j}_{i,\ell}v_{j}
\end{split}
\label{pre:connection}
\end{align}
We also have
\begin{align*}
	\Gamma^{\ell}_{\ell,i}=\partial_{i}\ln \sqrt{\det\mathsf{g}}
\end{align*}
We refer to e.g. \cite[\S~9]{FraT2012} for more details on differential geometry background.

\subsection{Ito and Stratonovich stochastic differential equations}

The conditional expectation value of the increment of a test function $f$ specifies the scalar generator $\operatorname{L}$ of the diffusion:
\begin{align*}
	(\operatorname{L}f)(\bm{q})=\lim_{s\downarrow 0}\operatorname{E}\left(\frac{f(\bm{\mathscr{q}}\td{t+s})-f(\bm{\mathscr{q}}\td{t})}{s}\Big{|}\bm{\mathscr{q}}\td{t}=\bm{q}\right)
	=b^{i}(\bm{q})\partial_{i}f(\bm{q})+\frac{1}{2}\mathsf{D}^{i,j}(\bm{q})\partial_{i}\partial_{j}f(\bm{q})
\end{align*} 
Using (\ref{pre:connection}), we derive the invariant expression of the generator:
\begin{align}
		(\operatorname{L}f)(\bm{q})=h^{i}(\bm{q})\nabla_{i}f(\bm{q})+\frac{1}{2}\mathsf{D}^{i,j}(\bm{q})\nabla_{i}\nabla_{j}f(\bm{q})
		\label{pre:generator}
\end{align} 
where
\begin{align}
	h^{\ell}(\bm{q})=b^{\ell}(\bm{q})+\frac{1}{2}\mathsf{D}^{i,j}(\bm{q})\Gamma_{i,j}^{\ell}(\bm{q})
\label{pre:Itodrift}
\end{align}
A direct calculation shows that $\bm{h}$ transforms under change of coordinates as a contravariant vector field.
The It\^o stochastic differential equation associated with (\ref{pre:generator}), however,  explicitly depends on the Christoffel symbols:
\begin{align}
	\mathrm{d}\mathscr{q}^{\ell}\td{t}=\left(h^{\ell}(\bm{\mathscr{q}}\td{t})-\frac{1}{2}\mathsf{D}^{i,j}(\bm{\mathscr{q}}\td{t})\Gamma_{i,j}^{\ell}(\bm{\mathscr{q}}\td{t})\right)\mathrm{d}t+\mathrm{e}_{i}^{\ell}(\bm{\mathscr{q}}\td{t})\mathrm{d}\mathscr{w}^{i}\td{t}
\label{pre:Ito}
\end{align}
 The equivalent Stratonovich representation
\begin{align}
	\mathrm{d}\mathscr{q}^{\ell}\td{t}=\left(h^{\ell}(\bm{\mathscr{q}}\td{t})-\frac{1}{2}\mathsf{D}^{i,j}(\bm{\mathscr{q}}\td{t})\left(\Gamma_{i,j}^{\ell}(\bm{\mathscr{q}}\td{t})-\tilde{\Gamma}_{i,j}^{\ell}(\bm{\mathscr{q}}\td{t})\right)\right)\mathrm{d}t+\mathrm{e}_{i}^{\ell}(\bm{\mathscr{q}}\td{t})\diamond\mathrm{d}\mathscr{w}^{i}\td{t}
	\label{pre:S}
\end{align}
 depends in addition on the {\textquotedblleft}spin{\textquotedblright} connection symbol:
\begin{align*}
\tilde{\Gamma}_{i,j}^{\ell}(\bm{q}):=	\sum_{k=1}^{d}\mathrm{e}_{k}^{\ell}(\bm{q})\partial_{i} (\tilde{\mathrm{e}}_{k})_{j}(\bm{q})
\end{align*}
The round brackets emphasize that the spin symbol $\tilde{\Gamma}_{i,j}^{\ell}$ depends on the $j$-th component of the elements of the covariant frame $\big{\{}\tilde{\bm{\mathrm{e}}}_{k}\big{\}}_{k=1}^{d}$ defined by (\ref{pre:cf}).
Only in the special case of {\textquotedblleft}integrable{\textquotedblright}  state-dependent noise \cite{ItoH1978}
\begin{align}
&	\tilde{\Gamma}_{i,j}^{\ell}(\bm{q})=\Gamma_{i,j}^{\ell}(\bm{q}) && \forall\,\bm{q}\,\in\,\mathbb{R}^{d}
\label{pre:integrable}
\end{align}
the Stratonovich drift coincides with the contravariant vector field appearing in the generator (\ref{pre:generator}).
A straightforward computation shows that this case is equivalently characterized by
\begin{align*}
&	\nabla_{i}\bm{\mathrm{e}}_{j}(\bm{q})=0 &&\forall\,\bm{q}\,\in\,\mathbb{R}^{d}
\end{align*}
which is sufficient but not necessary to satisfy the metric compatibility condition (\ref{pre:mcompatibility}).

In general, the difference between the Stratonovich and $\bm{h}$ is also a contravariant vector field. The Stratonovich equation
(\ref{pre:S}) is, however, not invariant under local rotations of the Wiener process. A proof of this well-known fact is detailed in  \cite[\S~4]{ItoH1978}. For {\textquotedblleft}non-integrable{\textquotedblright} state-dependent noise, the closest we can come to a contravariant stochastic differential equation for (\ref{pre:generator}) is  \cite[Th V.4.2]{IkWa1989} by considering  the Stratonovich system:
\begin{align*}
	\begin{cases}
&	\mathrm{d}\mathscr{q}^{\ell}\td{t}=h^{\ell}(\bm{\mathscr{q}}\td{t})\mathrm{d}t+\sigma^{\ell}_{k}\td{t}\diamond\mathrm{d}\mathscr{w}^{k}\td{t}
\\
&\mathscr{q}^{\ell}\td{0}=\bm{q}
\end{cases}
\end{align*}
coupled to the parallel transported frame
\begin{equation}
	\label{pre:pt}
		\begin{cases}
	&\mathrm{d}\sigma_{k}^{\ell}\td{t}=-\Gamma^{\ell}_{i,j}(\bm{\mathscr{q}}\td{t})\sigma^{i}_{k}\td{t}\diamond\mathrm{d}\mathscr{q}^{j}\td{t}
	\\
	&\sigma_{i}^{\ell}\td{0}=\bm{\mathrm{e}}_{i}^{\ell}(\bm{q})
\end{cases}
\end{equation}
Indeed, metric compatibility ensures that the collection of stochastic processes $\big{\{}\bm{\sigma}_{i}\big{\}}_{i=1}^{d}$ parallel transported along $\bm{\mathscr{q}}$ constitutes an orthonormal frame with respect to the metric $\mathsf{g}$ evaluated along the path of the diffusion at any time $t\,>\,0$ if it enjoys such property at $t=0$.


\section{Belopol'skaya-Daletskii representation}
\label{sec:BP}

In \cite[\S~2.2]{BeDa1982} Belopol'skaya and Daletskii propose an alternative representation of the Ito stochastic differential equation (\ref{pre:Ito}).  The idea is to use the exponential map to construct infinitesimal increments of the diffusion.
To this end, we recall that the exponential map
\begin{align*}
	\bm{\Phi}(\bm{q},\bm{v}):=\exp_{\bm{q}}(\bm{v})
\end{align*}
is the snapshot at $t=1$ of the solution 
\begin{align}
	\bm{\gamma}\td{t}=\exp_{\bm{q}}(\bm{v}\,t)
	\label{BD:em}
\end{align}
of the geodesic equation
\begin{align}
	\begin{split}
		&	\ddot{\gamma}^{\ell}\td{t}+\Gamma^{\ell}_{\,i,j}(\bm{\gamma}\td{t})\dot{\gamma}^{i}\td{t}\,\dot{\gamma}^{j}\td{t}=0
		\\
		&\bm{\gamma}\td{0}=\bm{q}
		\\
		&\bm{\dot{\gamma}}\td{0}=\bm{v}
	\end{split}
	\label{BD:geodesic}
\end{align}
In general, the exponential map is defined in a neighborhood of the point $\bm{q}$ (identified with its coordinates in an arbitrary frame of reference of $\mathbb{R}^{d}$) where the Peano-Picard  series converges. Furthermore, the representation (\ref{BD:em}) of the geodesic emphasizes that the natural expansion parameter of the Peano-Picard series is indeed the product $\bm{v}\,t$. In other words, for sufficiently {\textquotedblleft}small{\textquotedblright} $\bm{v}$ we can extricate the explicit approximate expression of the exponential map using the Peano-Picard series just by setting $t$ equal to one:
\begin{align}
	\Phi^{\ell}(\bm{q},\bm{v})=q^{\ell}+v^{\ell}-\frac{1}{2}\Gamma^{\ell}_{i,j}(\bm{q}) v^{i}\,v^{j}+O(\|\bm{v}\|)^{3}
	\label{BD:PP}
\end{align}
The gist of the Belopol'skaya-Daletskii representation \cite{BeDa1982} is that (\ref{pre:Ito})  generates the same path measure as
\begin{align}
	\mathrm{d}\bm{\mathscr{q}}\td{t}=\exp_{\bm{\mathscr{q}}\td{t}}\left(\bm{h}(\bm{\mathscr{q}}\td{t})\mathrm{d}t+\bm{\mathrm{e}}_{i}(\bm{\mathscr{q}}\td{t})\mathrm{d}\mathscr{w}^{i}\td{t}\right)-\bm{\mathscr{q}}\td{t}
	\label{BD:sde}
\end{align} 
We can heuristically verify the equivalence by combining (\ref{BD:PP}) with the finite deterministic value of the quadratic variation of the Wiener process:
\begin{align}
	\bm{\mathrm{e}}_{i}(\bm{\mathscr{q}}\td{t})\mathrm{d}\mathscr{w}^{i}\td{t}\,\bm{\mathrm{e}}_{j}(\bm{\mathscr{q}}\td{t})\mathrm{d}\mathscr{w}^{j}\td{t}= \mathsf{D}(\bm{\mathscr{q}}\td{t})\mathrm{d}t
	\label{BD:qv}
\end{align}
Notably, this identity also fixes the quadratic variation of the diffusion process
\begin{align*}
	\mathrm{d}\mathscr{q}^{i}\td{t}\mathrm{d}\mathscr{q}^{j}\td{t}=\mathsf{D}^{i,j}(\bm{\mathscr{q}}\td{t})\mathrm{d}t
\end{align*}
We emphasize that in (\ref{BD:sde}) we interpret the stochastic differential in the It\^o sense and that according to our conventions of volatility the following identity holds by definition
\begin{align*}
	\bm{\mathrm{e}}_{i}(\bm{\mathscr{q}}\td{t})\mathrm{d}\mathscr{w}^{i}\td{t}=\mathsf{A}(\bm{q})\mathrm{d}\bm{\mathscr{w}}\td{t}
\end{align*}
A statistically equivalent representation of the stochastic differential in (\ref{BD:sde}) is to replace the volatility with a path dependent frame evolving according to (\ref{pre:pt}) where now $\bm{\mathscr{q}}$ solves (\ref{BD:sde}).

\section{Finite dimensional approximations  to the covariant path measure}
\label{sec:pi}

The starting point is obtaining the short-time asymptotics $K^{\mathrm{s.t.}}$ of the transition probability density of the diffusion process $\bm{\mathscr{q}}$ over a time increment of length $ \varepsilon$. 

\subsection{Short time asymptotics}

We avail us of (\ref{BD:sde}) and of the volatility matrix introduced in (\ref{pre:volatitity})  to define the short-time approximation to the transition probability kernel:
\begin{align*}
	K^{(\mathrm{s.t.})}(\bm{y}_{i},t+\varepsilon\big{|}\bm{y}_{i-1},t)=\int_{\mathbb{R}^{d}}\mathrm{d}^{d}\bm{w}\frac{e^{-\frac{\left\|\bm{w}\right\|^{2}}{2\,\varepsilon}}}{(2\,\pi\,\varepsilon)^{d/2}}
	\delta^{(d)}\left(\bm{y}_{i}-\exp_{\bm{y}_{i-1}}\big{(}\bm{h}(\bm{y}_{i-1})\varepsilon+\mathsf{A}(\bm{y}_{i-1})\bm{w}\big{)}\right)
\end{align*}
The short-time kernel is readily positive and probability-preserving 
\begin{align*}
	\int_{\mathbb{R}}^{}\mathrm{d}^{d}\bm{y}\,K^{\mathrm{s.t.}}(\bm{y},t+\varepsilon\big{|}\bm{y}_{i-1},t)=1
\end{align*}
for every $\bm{y}_{i-1}\in\mathbb{R}^{d}$. For sufficiently small $\varepsilon$, only subsets of a neighborhood of $\bm{y}_{i-1}$ whose size vanish with $\varepsilon$  have a non-negligible probability. For such $\varepsilon$, the exponential map 
\begin{align*}
	\bm{x}=\exp_{\bm{q}}(\bm{v})
\end{align*} 
always admits an inverse:
\begin{align*}
	\bm{v}=\exp_{\bm{q}}^{-1}(\bm{x})
\end{align*}
We can therefore evaluate the integral over the Dirac-$\delta$ and obtain
\begin{align}
	K^{(\mathrm{s.t.})}(\bm{y}_{i},t+\varepsilon\big{|}\bm{y}_{i-1},t)=\frac{\sqrt{\det \mathsf{g}(\bm{y}_{i-1})}\,e^{-\frac{\mathcal{A}(\bm{y}_{j},\bm{y}_{j-1})}{2\,\varepsilon}}}{(2\,\pi\,\varepsilon)^{d/2}\det \mathsf{T}(\bm{y}_{i}|\bm{y}_{i-1})}
	\label{pi:st}
\end{align}
where
\begin{align*}
	\mathcal{A}(\bm{y}_{j},\bm{y}_{j-1}):=
	\left \langle\,\exp_{\bm{y}_{i-1}}^{-1}(\bm{y}_{i})-\bm{h}(\bm{y}_{i-1})\,\varepsilon\,,\exp_{\bm{y}_{i-1}}^{-1}(\bm{y}_{i})-\bm{h}(\bm{y}_{i-1})\,\varepsilon\,\right\rangle_{\mathsf{g}(\bm{y}_{i-1})}
\end{align*}
and
\begin{align*}
	\mathsf{T}(\bm{y}|\bm{q})=\bm{\partial}_{\bm{v}}\otimes\exp_{\bm{q}}(\bm{v})\big{|}_{\bm{v}=\exp_{\bm{q}}^{-1}(\bm{y})}
\end{align*}
Several remarks concerning (\ref{pi:st}) are in order before we proceed.
\begin{enumerate}[style=unboxed,leftmargin=0cm,label={\upshape\bfseries R-\roman*}]
	\item \label{R-i}The scalar transition density $S$ of a diffusion on a Riemann manifold must be normalized to unity with respect to the invariant volume element
	\begin{align}
		\operatorname{dvol}_{\mathsf{g}}(\bm{y})=\mathrm{d}^{d}\bm{y}\sqrt{\det\mathsf{g}(\bm{y})}
		\label{pi:vel}
	\end{align}
	Hence, (\ref{pi:st}) yields the following expression for the short-time asymptotics of the scalar density:
	\begin{align*}
		S^{\mathrm{s.t.}}(\bm{y}_{i},t+\varepsilon\big{|}\bm{y}_{i-1},t)=\frac{K^{\mathrm{s.t.}}(\bm{y}_{i},t+\varepsilon\big{|}\bm{y}_{i-1},t)}{\sqrt{\det\mathsf{g}(\bm{y}_{i})}}
	\end{align*}
	\item \label{R-ii} In order to derive (\ref{pi:st}), we took advantage of the fact that the volatility is non-singular (by hypothesis) and of the identities
	\begin{align*}
& \det\mathsf{A}(\bm{q})=\det\mathsf{A}^{\top}(\bm{q})=\sqrt{\det\mathsf{D}(\bm{q})}=\frac{1}{\sqrt{\det\mathsf{g}(\bm{q})}} && \forall\,\bm{q}\,\in\,\mathbb{R}^{d}
	\end{align*}
	\item \label{R-iii} The total {\textquotedblleft}Jacobian{\textquotedblright} prefactor generated by averaging the $\delta$-Dirac with respect to the Wiener process is
	\begin{align}
		J(\bm{y}_{i}|\bm{y}_{i-1})=\frac{\sqrt{\det \mathsf{g}(\bm{y}_{i-1})}}{(2\,\pi\,\varepsilon)^{d/2}\sqrt{\det \mathsf{g}(\bm{y}_{i})}\,\det \mathsf{T}(\bm{y}_{i}|\bm{y}_{i-1})}
		\label{BD:J}
	\end{align}
	\item \label{R-iv} At leading order, the Peano-Picard series yields 
	\begin{align}
	\big{(}\exp_{\bm{y}_{i-1}}^{-1}(\bm{y}_{i})\big{)}^{\ell}=y_{i}^{\ell}-	y_{i-1}^{\ell}+\frac{1}{2}\Gamma^{\ell}_{m,n}(\bm{y}_{i-1}) (y_{i}^{m}-	y_{i-1}^{m})(y_{i}^{n}-	y_{i-1}^{n})+\dots
	\label{pi:inverse}
	\end{align}
	Within this accuracy, we are in a position to recover the expressions of the argument of the exponential and Jacobian prefactor entering the short-time asymptotics obtained in \cite{GraR1985} from (\ref{pre:Ito}). Specifically,  if we approximate
	\begin{align}
		(\bm{y}_{i}-	\bm{y}_{i-1})\otimes(\bm{y}_{i}-	\bm{y}_{i-1})\approx \mathsf{D}(\bm{y}_{i-1})\,\varepsilon
		\label{pi:critical}
	\end{align}
	then we find
	\begin{align*}
		\mathcal{A}(\bm{y}_{i},\bm{y}_{i-1}) \approx 
		\left \langle\,\bm{y}_{i}-\bm{y}_{i-1}-\bm{b}(\bm{y}_{i-1})\varepsilon\,,\bm{y}_{i}-\bm{y}_{i-1}-\bm{b}(\bm{y}_{i-1})\varepsilon\,\right\rangle_{\mathsf{g}(\bm{y}_{i-1})}
	\end{align*}
	where $\bm{b}$ is the drift appearing in the It\^o stochastic differential equation (\ref{pre:Ito}) via (\ref{pre:Itodrift}). The recovery of the result in \cite{GraR1985} also requires 
	\begin{align*}
		J(\bm{y}_{i}|\bm{y}_{i-1})\approx \frac{\sqrt{\det \mathsf{g}(\bm{y}_{i-1})}}{(2\,\pi\,\varepsilon)^{d/2}\sqrt{\det \mathsf{g}(\bm{y}_{i})}}
	\end{align*}
	The approximation (\ref{pi:critical}) stems from the finite quadratic variation of the state-dependent noise (\ref{BD:qv}).  We already used this property of the quadratic variation to uphold the statistical equivalence of the Belopol'skaya-Daletskii equation (\ref{BD:sde}) with the It\^o form (\ref{pre:Ito}) of the stochastic differential equation associated with (\ref{pre:generator}). We emphasize that the short-time asymptotics coming from (\ref{pre:Ito}) retains only the zero-order contribution from $\mathsf{T}$:
	\begin{align*}
		\mathsf{T}(\bm{y}_{i}|\bm{y}_{i-1})\approx \mathsf{1}_{d}
	\end{align*}
	This approximation is the primary source of qualitative discrepancy between the results of \cite{GraR1985} and \cite{AnDr1999}. We expect, however, this discrepancy to become negligible when passing to the scaling limit that defines the path integral from finite-dimensional approximations. We revisit this point in section~\ref{sec:exe}.
\end{enumerate}

\subsection{Finite dimensional approximation of the path integral}

Let $ \mathcal{P}_{N}$ be a time lattice over $[0,t]$ with mesh $\varepsilon$ so that
\begin{align*}
	t=(N+1)\varepsilon
\end{align*}
As $N$ increases for fixed $t$, we generate finer and finer partitions of the interval. Correspondingly, the (strong) Markov property \cite[\S~V]{IkWa1989} allows us to uphold the scaling limit
\begin{align*}
	S(\bm{q},t|\bm{q}_{0},0)=\lim_{N\uparrow \infty}\frac{\operatorname{K}^{(\mathcal{P}_{N})}(\bm{q},t|\bm{q}_{0},0)}{\sqrt{\det \mathsf{g}(\bm{q})}}
\end{align*}
We recall that mathematically rigorous proofs of the existence of this limit over families of refining finite-dimensional approximations are the main result of \cite{AnDr1999} and \cite{BaPf2008}. In our heuristic treatment, finite-dimensional approximations read
\begin{align}
	\label{pi:fda}
		\frac{\operatorname{K}^{(\mathcal{P}_{N})}(\bm{q},t|\bm{y}_{0},0)}{\sqrt{\det \mathsf{g}(\bm{q})}}=\int 
		J(\bm{q}|\bm{y}_{N})\,e^{-\mathcal{A}(\bm{q},\bm{y}_{N})}\prod_{j=1}^{N}J(\bm{y}_{j}|\bm{y}_{j-1})
		\,e^{-\mathcal{A}(\bm{y}_{j},\bm{y}_{j-1})}\, \operatorname{dvol}_{\mathsf{g}}(\bm{y}_{j})
\end{align}
If we take into account remark~\ref{R-iv} above, (\ref{pi:fda}) recovers the expression of the finite-dimensional approximation 
to the path integral found in \cite{GraR1985}. To connect this result to \textbf{Theorem~1.8} of \cite{AnDr1999}, we need to further analyze  the Jacobian (\ref{BD:J}) in the large $N$ limit.

\subsection{Expression of the Jacobian}

The geometric meaning of the Jacobian (\ref{BD:J}) emerges if we relate it to the geodesic flow. Our aim is to present a heuristic yet logically controlled argument showing that as $N$ tends to infinity, in the finite-dimensional approximation (\ref{pi:fda}) we can 
write
\begin{align}
	J(\bm{y}_{j}|\bm{y}_{j-1})\overset{N\gg1}{\simeq}\frac{e^{\frac{R(\bm{y}_{j-1})\varepsilon_{}}{6}}}{(2\,\pi\,\varepsilon)^{d/2}}
	\label{pi:curvature}
\end{align}
with $R$ the curvature scalar of the metric $\mathsf{g}$. To this end, we observe that 
\begin{align*}
	\bm{\mathscr{v}}_{i}\td{t}=\partial_{v^{i}}\exp_{\bm{q}}(\bm{v}\,t)
\end{align*}
describes the Jacobi fields of the geodesic (\ref{BD:geodesic}) evolving from $\bm{q}$ generated by an infinitesimal variation of the $i$-th component of the initial velocity. A well-known result in Riemann geometry see, e.g., \cite[\S~10.1c]{FraT2012} shows that
the matrix whose columns are the linearly independent Jacobi fields $\bm{v}_{i}$
\begin{align*}
	\mathsf{V}\td{t}=\begin{bmatrix}
		\bm{\mathscr{v}}_{1}\td{t} & \dots &   \bm{ \mathscr{v}}_{d}\td{t} 
	\end{bmatrix}
\end{align*}
geometrically describes a $(1,1)$-tensor which obeys the linear equation
\begin{align}
	\begin{split}
		&	\frac{\nabla^{2}}{\mathrm{d}t^{2}}\mathsf{V}_{i}^{\ell}\td{t}+\operatorname{R}^{\ell}_{k,m,n}(\bm{\gamma}\td{t})\mathsf{V}_{i}^{m}\td{t}\,\dot{\mathscr{\gamma}}^{n}\td{t}\,\dot{\mathscr{\gamma}}^{k}\td{t}=0
		\\
		&\mathsf{V}\td{0}=0
		\\
		& 
		\dot{\mathsf{V}}\td{0}=\mathsf{1}_{d}
	\end{split}
\label{pi:Jacobi}
\end{align}
In the Jacobi equation (\ref{pi:Jacobi}) $\frac{\nabla}{\mathrm{d}t}$ is the covariant derivative along the geodesic $\bm{\gamma}$ which acts on any contravariant time-dependent vector as
\begin{align*}
	\frac{\nabla}{\mathrm{d}t} \mathscr{x}^{\ell}\td{t}:=\dot{\mathscr{x}}^{\ell}\td{t}+\Gamma^{\ell}_{i,j}(\bm{\gamma}\td{t})\mathscr{x}^{i}\td{t}\dot{\mathscr{\gamma}}^{j}\td{t}
\end{align*}
and $\mathsf{R}$ is the Riemann–Christoffel curvature tensor 
\begin{align*}
&\operatorname{R}^{\ell}_{i,m,n}:=\partial_{m}\Gamma^{\ell}_{i,n}-\partial_{n}\Gamma^{\ell}_{i,m}+\Gamma^{\ell}_{m,k}\Gamma^{k}_{i,n}-\Gamma^{\ell}_{n,k}\Gamma^{k}_{i,m}
\end{align*}
see, e.g., \cite[\S~8.5]{FraT2012} for further details.

In order to construct the solution of the Jacobi equation (\ref{pi:Jacobi}) we avail us of its relation with the flow solution of parallel transport along the same geodesic:
\begin{align}
		\begin{split}
		&	\frac{\nabla}{\mathrm{d}t}\mathsf{P}\td{t}=0
		\\
		& \mathsf{P}\td{0}=\mathsf{1}_{d}
	\end{split}
	\label{pi:pt}
\end{align}
We refer to $\mathsf{P}\td{t}$ as the parallel propagator.  Namely, upon resorting to the factorization Ansatz
\begin{align*}
	\mathsf{V}\td{t}=\mathsf{P}\td{t}\,\mathscr{V}\td{t}
\end{align*}
we arrive at
\begin{align*}
	&	\ddot{\mathscr{V}}_{\hspace{0.2cm}i}^{\ell}\td{t}+(\mathsf{P}^{-1}\td{t}\operatorname{R}\mathsf{P}^{-1}\td{t})^{\ell}_{k,m,n}(\bm{\gamma}\td{t})\mathscr{V}_{\hspace{0.2cm}i}^{m}\td{t}\dot{\mathscr{\gamma}}^{n}\td{t} \,\dot{\mathscr{\gamma}}^{k}\td{t}=0
	\\
	&\mathscr{V}\td{0}=0
	\\
	& \dot{\mathscr{V}}\td{0}=\mathsf{1}_{d}	
\end{align*}
Within first order accuracy in the Peano-Picard iteration, we obtain the solution
\begin{align*}
	\mathscr{V}^{\ell}_{\hspace{0.2cm}i}\td{t}=\delta_{\ell,i}\,t-\frac{t^{3}}{6}\operatorname{R}^{\ell}_{k,i,n}(\bm{q})\,v^{n}\,v^{k}+O(\|\bm{v}\|t)^{3}
\end{align*}
Next, we exploit the geometric properties of the parallel propagator $\mathsf{P}\td{t}$ solution of (\ref{pi:pt}) to couch it into a more transparent form. To this end, we recall that  any vector $\bm{v}$ in the tangent space of the starting point $\bm{q}$ of the geodesic, can be expressed in coordinates  specified by the orthogonal frame specified by eigenvectors of the diffusion tensor at $\bm{q}$:
\begin{align}
	\bm{v}=\bm{\mathrm{e}}_{i}(\bm{q})\xi^{i}\,\equiv\,\mathsf{A}(\bm{q})\bm{\xi}
	\label{pi:normal}
\end{align}
Parallel transport evolves elements of a frame along a curve by preserving the orientation. Hence we get
\begin{align*}
	\mathsf{P}\td{t}\bm{v}=\bm{\sigma}_{i}\td{t}\xi^{i}
\end{align*}
where now
\begin{align*}
			\begin{split}
		&	\frac{\nabla}{\mathrm{d}t}\bm{\sigma}_{i}\td{t}=0
		\\
		& \bm{\sigma}_{i}\td{0}=\bm{\mathrm{e}}_{i}(\bm{q})
	\end{split}
\end{align*}
Thus, we verify that the parallel propagator takes the form \cite[\S~5.2]{PoPoVe2011}
\begin{align}
	\mathsf{P}\td{t}=\sum_{i=1}^{d}\bm{\sigma}_{i}\td{t}\bm{\tilde{\mathrm{e}}}_{i}(\bm{q})
	\label{pi:pp}
\end{align}
The final step is to use the above expressions to evaluate
\begin{align*}
	\det \mathsf{T}(\bm{y}_{j}|\bm{y}_{j-1})&=\det\left( \mathsf{P}\td{1}\mathscr{V}\td{1}\big{|}_{\bm{v}_{j-1}=\exp_{\bm{y}_{j-1}}^{-1}(\bm{y}_{j})}\right)
\end{align*}
which factorizes into the product of two determinants that we can separately evaluate.
Within accuracy, upon recalling remark~\ref{R-ii}, we obtain 
\begin{align*}
	\det \left(\mathsf{P}\td{1}\big{|}_{\bm{v}_{j-1}=\exp_{\bm{y}_{j-1}}^{-1}(\bm{y}_{j})}\right)=\frac{\sqrt{\det \mathsf{g}(\bm{y}_{j-1})} }{\sqrt{\det \mathsf{g}(\bm{y}_{j})}}
\end{align*}
while by (\ref{pi:inverse}) and (\ref{pi:critical}) the second determinant yields
\begin{align*}
	\det\left(\mathscr{V}\td{1}\big{|}_{\bm{v}_{j-1}=\exp_{\bm{y}_{j-1}}^{-1}(\bm{y}_{j})}\right)=e^{\operatorname{Tr}\ln \left(\mathsf{1}_{d}-\sum_{i=1}^{d}\frac{\varepsilon}{6}\operatorname{R}^{\ell}_{k,\ell,n}(\bm{y}_{j-1})\mathrm{e}_{i}^{n}\mathrm{e}_{i}^{k}\right)}\overset{N\gg1}{\simeq} e^{-\frac{R(\bm{y}_{j-1})\,\varepsilon}{6}}
\end{align*}
Gathering all factors together we get
\begin{align*}
		\det \mathsf{T}(\bm{y}_{j}|\bm{y}_{j-1})\overset{N\gg1}{\simeq}\frac{\sqrt{\det \mathsf{g}(\bm{y}_{j-1})} }{\sqrt{\det \mathsf{g}(\bm{y}_{j})}}e^{-\frac{R(\bm{y}_{j-1})\,\varepsilon}{6}}
\end{align*}
Once we insert this result into (\ref{BD:J}) we finally obtain the announced result (\ref{pi:curvature}).

\subsection{Summary}

The upshot of the above considerations is that in the large $N$ limit  finite-dimensional approximations to the path measure generated by  the Belopol'skaya-Daletskii form of the stochastic differential equation with state-dependent noise take the form
\begin{align}
		S^{(\mathcal{P}_{N})}(\bm{q},t|\bm{y}_{0},0)\overset{N\gg1}{\simeq}\int 
	\,\frac{e^{-\mathcal{A}(\bm{q},\bm{y}_{N})+\frac{R(\bm{y}_{N})\varepsilon}{6}}}{(2\,\pi\,\varepsilon)^{d/2}}\prod_{j=1}^{N}
	\,e^{-\mathcal{A}(\bm{y}_{j},\bm{y}_{j-1})+\frac{R(\bm{y}_{j-1})\varepsilon}{6}}\, \frac{\operatorname{dvol}_{\mathsf{g}}(\bm{y}_{j})}{(2\,\pi\,\varepsilon)^{d/2}}
	\label{pi:AD}
\end{align}
as predicted by \textbf{Theorem~1.8}  of \cite{AnDr1999} and by \cite{BaPf2008}. We thus accomplished the task of  reconciling the elementary approach of \cite{GraR1985} with the more recent rigorous mathematical results of \cite{AnDr1999,BaPf2008}.
In particular, from (\ref{pi:AD}) we see that the prefactor $1/6$ of the scalar curvature is determined by the form (\ref{pi:vel}) 
of the volume element. This form is natural for any fixed finite-dimensional approximation. The reason to consider other definitions of the volume element in the scaling limit is that the infinite-dimensional Lebesgue measure is not well-defined. Hence, as already emphasized in \cite{LaRoTi1982} and proved in \cite{BaPf2008}, the scaling limit remains invariant if other definitions of the volume element are adopted, provided they are compensated by a judicious choice of the prefactor $c$ of the curvature scalar in the argument of the exponential.

\section{Examples of application}
\label{sec:exe}

One question remains open. The Belopol'skaya-Daletskii stochastic differential equation produces a short-time asymptotics of the transition probability density (\ref{pi:st}) that is non-Gaussian. This is at variance with asymptotics obtained from the It\^o stochastic differential equation (\ref{pre:Ito}). The question is how we recover from (\ref{pi:fda}) or equivalently (\ref{pi:AD}) the expressions of the scalar generator (\ref{pre:generator}) and, somewhat more interestingly, the  geodesic correction to stochastic parallel displacement of vector and tensor fields of \cite{DoGu1978,DoGu1979} referred to as the Weitzenb\"{o}ck formula in \cite[\S~V.5]{IkWa1989}.

\subsection{The role of normal coordinates}

The exponential map establishes a natural relation with Riemann normal coordinates \cite{MaMi1963}. Specifically, as long as there is a unique geodesic connecting $\bm{q}$ with any point in its neighborhood by continuously varying $\bm{v}$ in (\ref{BD:geodesic}), 
we can use 
\begin{align}
	\bm{x}=\exp_{\bm{q}}(\mathsf{A}(\bm{q})\bm{\xi})
	\label{exe:nc}
\end{align}
to uniquely associate the coordinates $\bm{x}$ of the tip of the geodesic starting from $\bm{q}$ with the coordinates of $\bm{\xi}$ specifying the initial velocity $\bm{v}$ of the geodesic  according to (\ref{pi:normal}). In this manner, we identify $\bm{\xi}$  with the so-called normal coordinates of the point represented by $\bm{x}$. The definition of the exponential map immediately implies that
\begin{align*}
	\bm{\gamma}\td{t}=\exp_{\bm{q}}(\mathsf{A}(\bm{q})\bm{\xi}\,t)
\end{align*}
satisfies (\ref{BD:geodesic}) so that 
\begin{align*}
	\bm{\gamma}^{(\mathrm{n.c.})}\td{t}=\bm{\xi}\,t
\end{align*}
is the geodesic in normal coordinates.

The above considerations motivate the change of variables in (\ref{pi:fda})
\begin{align*}
&	\bm{y}_{i}=\exp_{\bm{y}_{i-1}}(\bm{v}_{i-1})=\exp_{\bm{y}_{i-1}}(\mathsf{A}(\bm{y}_{i-1})\bm{\xi}_{i})
&& i=1,\dots,N
\end{align*} 
We imagine having solved these recursion relations to find
\begin{align*}
&	\bm{y}_{i}=\bm{\phi}_{i}(\bm{\xi}_{i},\dots,\bm{\xi}_{1},\bm{y}_{0}) && i=1,\dots,N
\end{align*}
with the convention that $\bm{\phi}_{0} $ is the identity map:
\begin{align*}
	\bm{y}_{0}=\bm{\phi}_{0}(\bm{y}_{0})\,\equiv\,\bm{q}_{0}
\end{align*}
Upon noticing that
\begin{align*}
	\mathsf{T}\left(\bm{\phi}_{i}|\bm{\phi}_{i-1}\right)\,=\,(\bm{\partial}_{\bm{v}_{i-1}}\otimes\exp_{\bm{\phi}_{i-1}})(\bm{v}_{i-1})\big{|}_{\bm{v}_{i-1}=\mathsf{A}(\bm{\phi}_{i-1})\bm{\xi}_{i}}
\end{align*}
it is straightforward to verify that the  Jacobian of the change of variables is a lower triangular matrix:
\begin{align*}
	&	\mathrm{d}\bm{y}_{1}=\mathsf{T}(\bm{\phi}_{1}|\bm{y}_{0})\mathsf{A}(\bm{y}_{0})\mathrm{d}\bm{\xi}_{1}
	\\
	&\mathrm{d}\bm{y}_{2}=\dots\mathrm{d}\bm{\xi}_{1}+\mathsf{T}(\bm{\phi}_{2}|\bm{\phi}_{1})\mathsf{A}(\bm{\phi}_{1})\mathrm{d}\bm{\xi}_{2}
	\\
	&\mathrm{d}\bm{y}_{3}=\dots\mathrm{d}\bm{\xi}_{1}+\dots\mathrm{d}\bm{\xi}_{2}+\mathsf{T}(\bm{\phi}_{3}|\bm{\phi}_{2})\mathsf{A}(\bm{\phi}_{2})\mathrm{d}\bm{\xi}_{3}
	\\
	&\vdots
\end{align*}
We indicate with dots non-diagonal entries of the Jacobian. Once we recall remark~\ref{R-ii}, we verify that under the change of variables the volume element changes as
\begin{align*}
	\prod_{i=1}^{N}\operatorname{dvol}_{\mathsf{g}}(\bm{y}_{i})=\prod_{i=1}^{N}\sqrt{\det\mathsf{g}(\bm{\phi}_{i})}\, \frac{\det\mathsf{T}(\bm{\phi}_{i}|\bm{\phi}_{i-1})}{\sqrt{\det\mathsf{g}(\bm{\phi}_{i-1})}}\mathrm{d}^{d}\xi_{i}
\end{align*}
The corresponding change of the integrand in (\ref{pi:fda}) is
\begin{align*}
	\prod_{i=1}^{N} J(\bm{\phi}_{i}|\bm{\phi}_{i-1})
	\,e^{-\mathcal{A}(\bm{\phi}_{i},\bm{\phi}_{i-1})}=\prod_{i=1}^{N}\frac{\sqrt{\det \mathsf{g}(\bm{\phi}_{i-1})}\,e^{-\mathcal{A}(\bm{\phi}_{i}|\bm{\phi}_{i-1})}}{(2\,\pi\,\varepsilon)^{d/2}\sqrt{\det \mathsf{g}(\bm{\phi}_{i})}\,\det \mathsf{T}(\bm{\phi}_{i}|\bm{\phi}_{i-1})}
\end{align*}
This means in  normal coordinates the finite-dimensional approximation reduces to
\begin{align*}
	&\operatorname{S}^{(\mathcal{P}_{N})}(\bm{q},t|\bm{q}_{0},0)=\int J(\bm{q}|\bm{\phi}_{N})\,e^{-\mathcal{A}(\bm{q},\bm{\phi}_{N})}e^{-\mathcal{A}^{(\mathrm{n.c.})}(\bm{\xi}_{N},\dots,\bm{\xi}_{1},\bm{q}_{0})} \prod_{i=1}^{N}\frac{\mathrm{d}^{d}\xi_{i}}{(2\,\pi\,\varepsilon)^{d/2}}
\end{align*}
with
\begin{align*}
	\mathcal{A}^{(\mathrm{n.c.})}(\bm{q},\bm{\xi}_{N},\dots,\bm{\xi}_{1},\bm{q}_{0})=\sum_{j=1}^{N}\left(\frac{\|\bm{\xi}_{j}\|^{2}}{2\,\varepsilon} -\left \langle\,\mathsf{A}(\bm{\phi}_{j-1})\bm{\xi}_{j}\,,\bm{h}(\bm{\phi}_{j-1})\,\right\rangle_{\mathsf{g}(\bm{\phi}_{j-1})}+\frac{\|\bm{h}(\bm{\phi}_{j-1})\|_{\mathsf{g}(\bm{\phi}_{j-1})}^{2}\,\varepsilon}{2}\right) 	
\end{align*}
The final step consists in using the definition of the volatility to reduce the inner product with respect to the Riemann metric to the standard Euclidean inner product:
\begin{align*}
	\left \langle\,\mathsf{A}(\bm{\phi}_{j-1})\bm{\xi}_{j}\,,\bm{h}(\bm{\phi}_{j-1})\,\right\rangle_{\mathsf{g}(\bm{\phi}_{j-1})}
	=\left \langle\,\bm{\xi}_{j}\,,\mathsf{A}^{-1}(\bm{\phi}_{j-1})\bm{h}(\bm{\phi}_{j-1})\,\right\rangle
\end{align*}
Upon recalling (\ref{pi:normal}) we write
\begin{align*}
	\bm{h}(\bm{\phi}_{j-1})=\mathsf{A}(\bm{\phi}_{j-1})\bm{h}^{(\mathrm{n.c.})}(\bm{\phi}_{j-1})
\end{align*}
and similarly
\begin{align*}
	\|\bm{h}(\bm{\phi}_{j-1})\|_{\mathsf{g}(\bm{\phi}_{j-1})}^{2}=\|\bm{h}^{(\mathrm{n.c.})}(\bm{\phi}_{j-1})\|^{2}
\end{align*}
where on the right hand side appears the Euclidean norm squared of the normal coordinate representation of the vector field $\bm{h}$.
The upshot is
\begin{align*}
	\mathcal{A}^{(\mathrm{n.c.})}(\bm{\xi}_{N},\dots,\bm{\xi}_{1},\bm{q}_{0})=\sum_{j=1}^{N}\frac{\| \bm{\xi}_{j}-\bm{h}^{(\mathrm{n.c.})}(\bm{\phi}_{j-1})\|^{2}}{2\,\varepsilon}
\end{align*}
This is exactly the result that we obtain if we evaluate in normal coordinates the finite-dimensional approximation derived by \cite{GraR1985} because in normal coordinates Christoffel symbols evaluated at the origin vanish.  Therefore, if we use normal coordinates to compute the scaling limit of finite-dimensional approximations the formula derived in \cite{GraR1985} and those of \cite{AnDr1999,BaPf2008} must certainly give the same result. 

To illustrate the self-consistency of the short-time asymptotics we apply it to recover two well-known results.

\subsection{Recovery of the scalar generator}
\label{sec:g}

We wish to compute the scalar generator of the diffusion from the limit 
\begin{align*}
	(\operatorname{L}f)(\bm{y})=\lim_{\varepsilon\downarrow 0}	\frac{1}{\varepsilon}\left(\int_{\mathbb{R}^{d}}\mathrm{d}^{d}q \,f(\bm{q}) \frac{\sqrt{\det \mathsf{g}(\bm{y})}}{(2\,\pi\,\varepsilon)^{d/2}}\,e^{-\frac{\|\exp_{\bm{y}}^{-1}(\bm{q})-\bm{h}(\bm{y})\,\varepsilon\|_{\mathsf{g}(y)}^{2}}{2\,\varepsilon}}-f(\bm{y})
	\right)
\end{align*}
Changing variables to the Riemann normal coordinates $\bm{\xi}$, we get 
\begin{align*}
		\operatorname{E}(f(\bm{\mathscr{q}}\td{\varepsilon})\big{|}\bm{\mathscr{q}}\td{0}=\bm{y})=\int \frac{\mathrm{d}^{d}\xi}{(2\,\pi\,\varepsilon)^{d/2}} \,f\left(\exp_{\bm{y}}(\mathsf{A}(\bm{y})\bm{\xi})\right)
	e^{-\frac{\|\bm{\xi}-\mathsf{A}^{-1}(\bm{y})\bm{h}(y) \,\varepsilon\|^{2}}{2\,\varepsilon}}
\end{align*}
The rescaling
\begin{align*}
	\bm{\xi}\mapsto\sqrt{\varepsilon}\,\bm{\xi}
\end{align*}
yields
\begin{align*}
	\operatorname{E}(f(\bm{\mathscr{q}}\td{\varepsilon})\big{|}\bm{\mathscr{q}}\td{0}=\bm{y})=\int \frac{\mathrm{d}^{d}\xi}{(2\,\pi)^{d/2}} \,f\left(\exp_{\bm{y}}(\sqrt{\varepsilon}\mathsf{A}(\bm{y})\bm{\xi})\right)
	e^{-\frac{\|\bm{\xi}-\mathsf{A}^{-1}(\bm{y})\bm{h}(y) \,\sqrt{\varepsilon}\|^{2}}{2}}
\end{align*}
To evaluate the limit we expand everything up to order $O(\varepsilon)$:
\begin{align*}
	&	f\left(\exp_{\bm{y}}(\sqrt{\varepsilon}\mathsf{A}(\bm{y})\bm{\xi})\right)
	=f(\bm{y})+\sqrt{\varepsilon}\left \langle\,\mathsf{A}(\bm{y})\bm{\xi}\,,(\bm{\partial}f)(\bm{y})\,\right\rangle
	\\
	&-\frac{\varepsilon}{2}\left \langle\,\Gamma( \mathsf{A}(\bm{y})\bm{\xi},\mathsf{A}(\bm{y})\bm{\xi})\,,(\bm{\partial}f)\left(\bm{y}\right)\,\right\rangle+\frac{\varepsilon}{2}\left \langle\,\mathsf{A}(\bm{y})\bm{\xi}\otimes\mathsf{A}(\bm{y})\bm{\xi} \,,(\bm{\partial}\otimes\bm{\partial}f)\left(\bm{y}\right)\,\right\rangle+O(\varepsilon^{3/2})
\end{align*}
where we use the abridged notation
\begin{align*}
	\Gamma^{\ell}(\bm{a},\bm{b})=\Gamma^{\ell}_{i,j}a^{i}b^{j}
\end{align*}
and
\begin{align*}
	&	e^{-\frac{\|\bm{\xi}-\mathsf{A}^{-1}(\bm{y})\bm{h}(y) \,\sqrt{\varepsilon}\|^{2}}{2}}=	e^{-\frac{\|\bm{\xi}\|^{2}}{2}}\left(1+\sqrt{\varepsilon}\left \langle\,\bm{\xi}\,,(\mathsf{A}^{-1}\bm{h})(\bm{y})\,\right\rangle\right)
	\\
	&\,-\varepsilon\,e^{-\frac{\|\bm{\xi}\|^{2}}{2}}\,\left(\frac{\|(\mathsf{A}^{-1}\bm{h})(\bm{y})\|^{2}-\left \langle\,\bm{\xi}\,,(\mathsf{A}^{-1}\bm{h})(\bm{y})\,\right\rangle\left \langle\,\bm{\xi}\,,(\mathsf{A}^{-1}\bm{h})(\bm{y})\,\right\rangle}{2}\right)+O(\varepsilon^{3/2})
\end{align*}
The integral over $\bm{\xi}$ is the average of a Gaussian vector with zero mean and unit variance. Upon recalling (\ref{pre:D}),  we readily recover the form of the generator.

\subsection{Recovery of Dohrn-Guerra stochastic geodesic displacement of vector fields}
\label{sec:DG}

We now turn our attention to the mean forward derivative of a vector field. As shown by Dohrn and Guerra \cite{DoGu1978,DoGu1979}, the evaluation of this quantity on a Riemann manifold requires taking into account the geodesic correction to stochastic parallel transport. In other words, if we denote by  $\mathsf{X}\td{t}$ the (1,1)-tensor of Jacobi fields  
along the geodesic $\bm{\gamma}$ connecting $\bm{q}$ to $\bm{\mathscr{q}}\td{t}$
\begin{align}
	\begin{split}
		&	\frac{\nabla^{2}}{\mathrm{d}t^{2}}\mathsf{X}_{i}^{\ell}\td{t}+\operatorname{R}^{\ell}_{k,m,n}(\bm{\gamma}\td{t})\mathsf{X}_{i}^{m}\td{t}\,\dot{\mathscr{\gamma}}^{n}\td{t}\,\dot{\mathscr{\gamma}}^{k}\td{t}=0
		\\
		&\mathsf{X}\td{0}=\operatorname{1}_{d}
		\\
		& \left (\frac{\nabla}{\mathrm{d}t}\mathsf{X}\right )^{\ell}_{i}\td{0}\,\equiv\,\dot{\mathsf{X}}^{\ell}_{i}\td{0}+\Gamma^{\ell}_{m,n}(\bm{\gamma}\td{0})\mathsf{X}_{k}^{m}\td{0}\,\dot{\mathscr{\gamma}}^{n}\td{0}=0
	\end{split}
	\label{exe:Jacobi}
\end{align}
describing the evolution of an initial perturbation of the starting point of the geodesic while holding fixed the initial velocity, then a consistent definition of the mean forward derivative of a smooth vector field $\bm{f}$ along the diffusion $\bm{\mathscr{q}}$ is \cite{DoGu1978,DoGu1979}:
\begin{align*}
	\operatorname{D}_{+}\bm{f}(\bm{y})=\lim_{t\downarrow 0}\frac{\operatorname{E}
	\left( \mathsf{X}^{-1}\td{t}\bm{f}(\bm{\mathscr{q}}\td{t})-\bm{f}(\bm{\mathscr{q}}\td{0})\big{|}\bm{\mathscr{q}}\td{0}=\bm{y}\right)}{t}
\end{align*}
To gain geometric insight into the definition, we resort again to the relation between parallel transport and Jacobi fields by writing
\begin{align*}
	\mathsf{X}\td{t}=\mathsf{P}\td{t}\,\mathscr{X}\td{t}
\end{align*}
In this case, we find
\begin{align*}
		&	\ddot{\mathscr{X}}_{\hspace{0.2cm}i}^{\ell}\td{t}+(\mathsf{P}\td{t}^{-1}\operatorname{R}\mathsf{P}\td{t}^{-1})^{\ell}_{k,m,n}(\bm{\gamma}\td{t})\mathscr{X}_{\hspace{0.2cm}i}^{m}\td{t}\dot{\mathscr{\gamma}}^{n}\td{t} \,\dot{\mathscr{\gamma}}^{k}\td{t}=0
	\\
	&\mathscr{X}\td{0}=\mathsf{1}_{d}	
	\\
	& \dot{\mathscr{X}}\td{0}=0
\end{align*}
At leading order in the Peano-Picard series,  we obtain
\begin{align*}
	\mathscr{X}^{\ell}_{\hspace{0.2cm}i}\td{t}=\delta_{\ell,i}-\frac{t^{2}}{2}\operatorname{R}^{\ell}_{k,i,n}(\bm{q})v^{n}v^{k}+O(\|\bm{v}\|\,t)^{3}
\end{align*}
Once we recall the representations (\ref{pi:pp}) of the parallel propagator  and of vector fields in terms of frames
\begin{align*}
\bm{f}(\bm{\mathscr{q}}\td{t})=\sum_{i=1}^{d} \bm{\sigma}_{i}\td{t}\mathscr{f}_{i}(\bm{\mathscr{q}}\td{t})	
\end{align*}
we can couch the mean forward derivative into the form
\begin{align*}
	\operatorname{D}_{+}\bm{f}(\bm{y})=\lim_{t\downarrow 0}\operatorname{E}
		\left( \sum_{i=1}^{d}\frac{\mathscr{X}^{-1}\td{t} \bm{\sigma}_{i}\td{0}\mathscr{f}_{i}(\bm{\mathscr{q}}\td{t})-\bm{\sigma}_{i}\td{0}\mathscr{f}_{i}(\bm{\mathscr{q}}\td{0})}{t}\Big{|}\bm{\mathscr{q}}\td{0}=\bm{y}\right)
\end{align*}
This last expression exhibits the geodesic correction to stochastic parallel transport. In addition, the $\mathscr{f}_{i}(\bm{\mathscr{q}}\td{t})$ geometrically behave as scalars. We thus expect
\begin{align}
	\operatorname{D}_{+}f^{\ell}(\bm{y})=h^{i}(\bm{q})\nabla_{i}f^{\ell}(\bm{q})+\frac{1}{2}\mathsf{D}^{i,j}(\bm{q})\nabla_{i}\nabla_{j}f^{\ell}(\bm{q})+\frac{1}{2}\operatorname{Ric}^{\ell}_{i}(\bm{q})f^{i}(\bm{q})
	\label{exe:DG}
\end{align}
where the Ricci tensor is
\begin{align*}
	\operatorname{Ric}^{\ell}_{i}(\bm{q})=\operatorname{R}^{\ell}_{m,i,n}(\bm{q})\mathsf{D}^{m,n}(\bm{q})
\end{align*}
We can check the correctness of the above geometric argument, by repeating the above calculations in arbitrary coordinates.
In such a case, we get
\begin{align*}
(\mathsf{X}^{-1})^{\ell}_{k}\td{t}&=\delta_{\ell,k}+\Gamma^{\ell}_{k,i}(\bm{q})v^{i}\,t
\nonumber\\
&+	\frac{v^{i}\,v^{j}\,t^{2}}{2}\left(\operatorname{R}^{\ell}_{i,k,j}(\bm{q})
	-\Gamma^{\ell}_{k,m}(\bm{q})\Gamma^{m}_{i,j}(\bm{q})
	+\Gamma^{\ell}_{i,m}(\bm{q})\Gamma^{m}_{k,j}(\bm{q})
	+(\partial_{i}\Gamma^{\ell}_{k,j})(\bm{q})\right )+O(\|\bm{v}\|t)^{3}
\end{align*}
and we need to recast the Taylor expansion of the vector field $\bm{f}$ in terms of its covariant derivatives. The calculation is
laborious as $\nabla_{i}\bm{f}$ is now a (1,1)-tensor. Hence, the expression of the generator in terms of covariant derivatives requires the identity:
\begin{align*}
	 \nabla_{j}\nabla_{i}f^{\ell}(\bm{q})
	=\partial_{j}\nabla_{i}\, f^{\ell}(\bm{q})
	+\Gamma^{\ell}_{j,m}(\bm{q})\nabla_{i}f^{m}(\bm{q})-\Gamma^{m}_{i,j}(\bm{q})\nabla_{m}f^{\ell}(\bm{q})
\end{align*}
Bearing this information in mind together with some algebra to patiently perform all the necessary cancellations  yields (\ref{exe:DG}), as expected.

\section{Conclusions}

The construction of finite-dimensional approximations of the path measure  of a diffusion from the stochastic differential equation
directly connects analytic arguments with Monte Carlo methods used in numerical approaches. This is probably the main reason for the periodically resurfacing interest not only in the physics; see, e.g., \cite{LaRoTi1982,CuLe2017,PiCuLeWi2023} but also in the mathematical literature, see \cite{BrGaHaZa2021} in investigating the origin of the stumbling blocks that one encounters when trying to derive the expression of the path measure in terms of scalar quantities from the familiar It\^o, Stratonovich or Klimontovich representations of the stochastic differential equations on Riemann manifolds. In this note we show that the less-known Belopol'skaya-Daletskii formulation of stochastic differential equations on Riemann manifolds  \cite{BeDa1982,BeDa1990} avoids the obstacles encountered by other approaches. 
The gist of the Belopol'skaya-Daletskii is to explicitly use the exponential map to construct stochastic increments. This idea is present but not fully explicit in the rigorous treatment \cite{AnDr1999,BaPf2008} and to some extent considered in the formulation of \textbf{Conjecture 4.8} of \cite{BrGaHaZa2021} (see  equation (59) therein).  These considerations suggest that explicitly using the Belopol'skaya-Daletskii  formalism may provide a welcome avenue to simplify existing proofs  in the finite-dimensional case and perhaps inspire further advances in the infinite-dimensional case considered in \cite{BrGaHaZa2021}. 

Finally,  given a metric we can envisage training a neural network to learn the exponential map on a Riemann manifold. If this can be done consistently and efficiently, then the learned exponential map provides a black box algorithm to generate the increments of a diffusion process on a Riemann manifold. This numerical strategy may offer advantages in terms of speed and scalability with respect to standard numerical approaches based on standard formulations of stochastic differential equations. This is a suggestion, however, whose concrete benefits require further analysis.

\section{Acknowledgments}

We gratefully acknowledge discussions with D. Lucente, M. Baldovin, L. Magazz\` u, and E.S. Ceruleo. 


\vspace{.2cm}
\section*{References}	

\bibliographystyle{iopart-num}
\bibliography{covariant_sde}{} 
\end{document}